\documentclass[aip,reprint]{revtex4-1}
\usepackage[space]{grffile}
\usepackage{bm,graphicx,graphics,amsmath,amssymb,bm,epsfig,color}
\usepackage{euscript,multirow,tabularx}
\usepackage{longtable}
\usepackage{float}
\usepackage[draft]{todonotes}  
\usepackage{pifont}
\usepackage{comment}

\graphicspath{{./figs/final/}}

\begin{document}

\title{Enhancement of YIG$|$Pt spin conductance by local Joule annealing}

\author{R. Kohno}
\affiliation{Universit\'e Grenoble Alpes, CEA, CNRS, Spintec, 38054 Grenoble, France}

\author{N. Thiery}
\affiliation{Universit\'e Grenoble Alpes, CEA, CNRS, Spintec, 38054 Grenoble, France}

\author{K. An}
\affiliation{Universit\'e Grenoble Alpes, CEA, CNRS, Spintec, 38054 Grenoble, France}

\author{P. Noel}
\affiliation{Universit\'e Grenoble Alpes, CEA, CNRS, Spintec, 38054 Grenoble, France}

\author{L. Vila}
\affiliation{Universit\'e Grenoble Alpes, CEA, CNRS, Spintec, 38054 Grenoble, France}

\author{V. V. Naletov} 
\affiliation{Universit\'e Grenoble Alpes, CEA, CNRS, Spintec, 38054 Grenoble, France}
\affiliation{Institute of Physics, Kazan Federal University, Kazan
    420008, Russian Federation}

\author{N. Beaulieu} 
\affiliation{SPEC, CEA-Saclay, CNRS, Universit\'e Paris-Saclay,
  91191 Gif-sur-Yvette, France}
\affiliation{LabSTICC, CNRS, Universit\'e de Bretagne Occidentale,
  29238 Brest, France}

\author{J. Ben Youssef} 
\affiliation{LabSTICC, CNRS, Universit\'e de Bretagne Occidentale,
  29238 Brest, France}

\author{G. de Loubens} 
\affiliation{SPEC, CEA-Saclay, CNRS, Universit\'e Paris-Saclay,
  91191 Gif-sur-Yvette, France}

\author{O. Klein}
\email[Corresponding author:]{ oklein@cea.fr}
\affiliation{Universit\'e Grenoble Alpes, CEA, CNRS, Spintec, 38054 Grenoble, France}

\date{\today}

\begin{abstract}
{We report that Joule heating can be used to enhance the interfacial spin conductivity between a metal and an oxide. We observe that local annealing of the interface at about 550\,K by injecting large current densities ($>10^{12}\text{A/m}^{2}$) into a pristine 7\,nm thick Pt nanostrip evaporated on top of yttrium iron garnet (YIG), can improve the spin transmission up to a factor 3: a result of particular interest for interfacing ultra thin garnet films where strong chemical etching of the surface has to be avoided. The effect is confirmed by different methods: spin Hall magnetoresistance, spin pumping and non-local spin transport. We use it to study the influence of the YIG$|$Pt coupling on the non-linear spin transport properties. We find that the cross-over current from a linear to a non-linear spin transport regime is independent of this coupling, suggesting that the behavior of pure spin currents circulating in the dielectric are mostly governed by the physical properties of the bare YIG film beside the Pt nanostrip.} 
\end{abstract}

\maketitle

The transport of pure spin information through localized magnetic moments is at the heart of a new research topic called \textit{insulatronic} (for \textit{insula}-tor spin-\textit{tronic}) \cite{cornelissen2015long,goennenwein2015non,kajiwara2010transmission,brataas2020}. 
Interest stems here from the recognition that magnetic insulators are superior spin conductors than metals or semi-conductors. Among magnetic insulators, garnets, and in particular yttrium iron garnet (YIG), have the lowest known magnetic damping \cite{cherepanov1993saga}. One can exploit here the spin Hall effect (SHE) to interconvert pure spin currents circulating inside the dielectric into charge currents, which can then be probed electrically. This is usually achieved by depositing a heavy metal electrode, advantageously in Pt \cite{valenzuela2006direct,sanchez2013spin}, on top of the YIG surface. The ensuing flow of spin escaping through the metal-oxide interface  can be measured through spin pumping (SP) \cite{Heinrich2003}, spin Hall magnetoresistance (SMR) \cite{hahn2013comparative,wang2017comparative,nakayama2013spin}, spin Seebeck effect (SSE) \cite{uchida2010observation} or spin orbit torque (SOT) using non-local transport devices \cite{cornelissen2015long,thiery2018nonlinear}. 

The efficiency of the process is determined by the spin transparency of the interface and parametrized by the so-called spin mixing conductance, $G_{\uparrow\!\downarrow}$. To optimize its strength, the YIG surface is usually treated by strong process such as O$^+$/Ar$^+$ plasma \cite{velez2016competing}, by annealing \cite{qiu2013spin}  and piranha etching\cite{jungfleisch2013improvement,putter2017impact} prior to the metal deposition in order to achieve good chemical and structural YIG$|$Pt interface. However these treatments are performed on $\mu$m thick YIG samples and are difficult to implement once the thickness of the film is in the nanometer range as a modulation of surface roughness significantly disturbs its magnetic properties. Additionally an enhancement of SMR by global annealing at very high temperature\cite{saiga2014platinum} has been reported, yet this process is not necessarily compatible at the device level. In addition to the cleaning of the YIG surface, the use of sputtering technique to deposit the metal is known to lead to better $G_{\uparrow\!\downarrow}$ than the evaporation technique\cite{vlietstra2013spin}. The later however usually leads to lower resistivity, which should be favored when one wants to inject large current densities. Considering that evaporation is also advantageous to have better lift-off during nanofabrication, certainly a process allowing to improve interfacial quality of evaporated Pt is required.\\

In this paper, we investigate the impact of local Joule heating at 550\,K on the spin transport between thin YIG film and evaporated Pt. We use SMR and spin pumping measurements to show a clear 3 times post-deposition enhancement of the interfacial spin transmission, which is irreversible. {We have exploited this feature to study the influence of the interfacial spin conductivity on non-linear spin transport properties in lateral devices. We find that the later are mostly governed by the physical properties of the bare YIG film not covered by Pt. Any enhancement of the coupling to the electrodes seems to play a negligible role in determining the non-linear characteristics of pure spin transport.}
 
\begin{figure}
\begin{center}
\includegraphics[width=8.4cm,height=6.6cm]{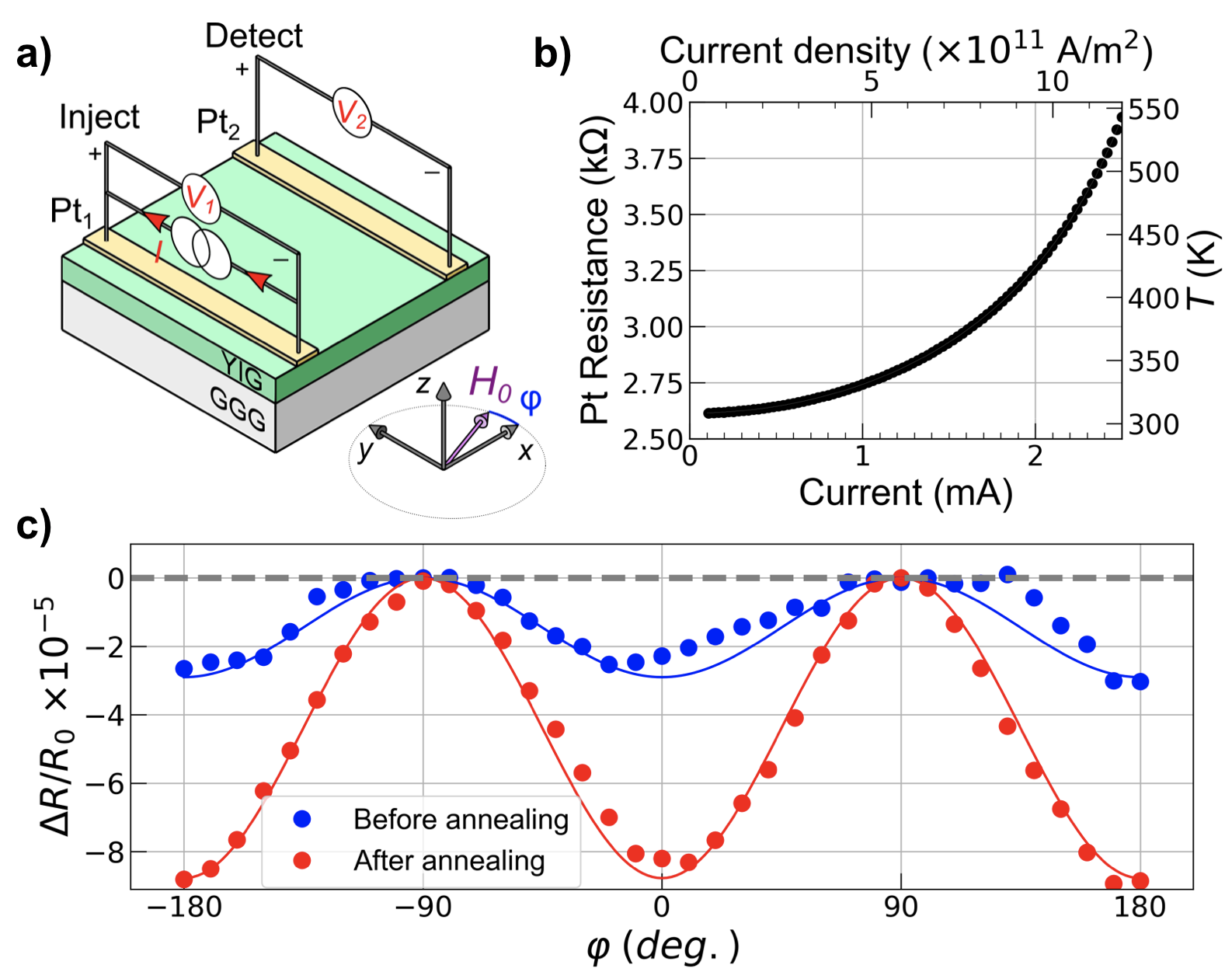}
\end{center}
\caption{
a) Schematic of the YIG$|$Pt structure with two nanostrips oriented along the $y$-direction. The first one (Pt$_1$) is connected to a current source and it is  used to measure the SMR ratio. The second electrode (Pt$_2$) is used to measure the non-local spin transport properties. b) Resistance of the Pt$_1$ nanostrip, $R_I=V_1/I$, as a function of the injected current $I$ inside. c) Angular dependence of the SMR ratio when an applied magnetic $\mu_0 H_{0}$= 200\,mT is rotated in the $xy$ plane. The data shows the result before (blue dots) and after (red dots) local Joule annealing at 550\,K. Solid lines are fit with $\cos^2\varphi$.
}
\label{Fig1}
\end{figure}

We use a $t_\text{YIG}=56$\,nm thick YIG film grown by liquid phase epitaxy on a 500\,$\mu$m Gd$_3$Ga$_5$O$_{12}$ (GGG) substrate \cite{hahn2013comparative,castel2012frequency}. {Ferromagnetic resonance experiments have shown a damping parameter of  $\alpha_\text{YIG} = 2.2\, \times 10 ^{-4}$ revealing an excellent crystal quality of the YIG film \cite{beaulieu2018temperature}.} Two similar Pt nanostrips, respectively Pt$_1$ and Pt$_2$, are patterned by e-beam lithography to have a width of 300\,nm and length of 30\,$\mu$m. A 7$\,$nm thick Pt layer is then deposited by e-beam evaporation on the YIG film. The nanostrips are connected to Ti$|$Au (5\,nm$|$50\,nm) electrical contacts. The sample is mounted on a rotational stage and exposed to an in-plane magnetic field of $\mu_0 H_{0}$= 200\,mT to fully magnetize the YIG film. All the magneto-transport experiments are performed at room temperature, $T_0$. The schematic of the sample geometry is shown in Fig.\ref{Fig1}a). Transport measurements are performed by injecting 10\,ms current pulses with a duty cycle of 10\% into the Pt$_1$ nanostrip via a 6221 Keithley current source which is synchronized to a 2182A Keithley nanovoltmeter. We first investigate the configuration where the nanovoltmeter is connected to the same nanostrip to extract the magnetoresistive response through $V_1$ \cite{thiery2018electrical}. Fig.\ref{Fig1}b) represents the evolution of  the Pt$_1$ resistance, $R_I$, as a function of the electrical current,  $I$, showing a quadratic dependence due to Joule heating. The Pt electrode can also be used as a temperature sensor. The temperature rise of the Pt is simply inferred from the change of Pt resistance with $T -T_0 = \zeta_{Pt}$($R_{I}$ - $R_{0}$)/$R_{0}$, where $R_0= 2.6$ k$\Omega$ is the Pt resistance at $I=0$ and $\zeta_{Pt}$ = 478\,K is a thermal coefficient specific to our Pt nanostrip. In our structure, the local temperature reaches about 550\,K when the current density is $J_\text{max} = 1.2 \times$ $10^{12}$ A/m$^{2}$.

The spin transparency of the interface can be evaluated from the SMR ratio defined as $(R_{\varphi} -  R _{\|}) / R_0$, which measures the relative change of the Pt resistance as a function of $\varphi$, the azimuthal angle between $H_0$ (and thus the magnetization) and the $x$-axis. In this notation $R _{\|}$ is the resistance when $H_0$ is applied parallel to the nanostrip direction ($\varphi=\pm90^\circ$ or $y$-axis). Fig.\ref{Fig1}c) presents the SMR signal measured with a bias current of $I=100\,\mu$A. The data in blue dots show the values obtained directly after the nanofabrication process. 
The angular dependence follows a $\cos^2\varphi$ behavior (see solid line fit), and the maximum SMR deviation is observed when $H_0$ is applied perpendicular to the nanostrip direction ($\varphi=0^\circ$ or $x$-axis) with a value of $(R_\perp - R_\|) / R_0 = -2.9\times10^{-5}$ extracted from the fit (blue line). 
From the theory of SMR\cite{chen2013theory,nakayama2013spin}, the amplitude of the SMR ratio is expressed as :
\begin{equation}
\dfrac{R_\perp - R_\|}{R_0} = -\theta_\text{SHE}^{2}\dfrac{2({\lambda_\text{{sf}}^{2}}/{t_\text{Pt}}) \rho g_{\uparrow\!\downarrow} \tanh^{2}\left (\dfrac{t_\text{Pt}}{2\lambda_\text{sf}}\right )}{1 + 2\lambda_\text{sf}\rho g_{\uparrow\!\downarrow} \coth\left (\dfrac{t_\text{Pt}}{\lambda_\text{sf}} \right )},
\end{equation}
where $\rho= 19.5\,\mu\Omega$cm is the Pt resistivity with $t_\text{Pt}$ its thickness, $\lambda_{\text{sf}}$ is  the spin diffusion length, and $\theta_{\text{SHE}}$ is the SHE angle inside the Pt layer. In our notation, $g_{\uparrow\!\downarrow}$ is an effective spin mixing conductance of the YIG$|$Pt interface (see discussion below).

Using SMR measurement, we then investigate the effect of 550\,K local Joule annealing on the effective spin mixing conductance of YIG$|$Pt. The electrical annealing is provided by applying current density pulses of $J_\text{max} = 1.2\,\times\,10^{12}$ A/m$^2$  {for about 60 minutes}.  
The red dots shows the SMR ratio measured after. The amplitude is increased to $(R_\perp - R_\|) / R_0 =-8.9 \times 10^{-5}$ (red line fit in Fig.\ref{Fig1}c), which is 3 times larger than the value before annealing  (blue line). This enhanced SMR is irreversible and increasing the annealing time above an hour leads to negligible gain in the SMR value. We also observe that the Pt$_1$ resistance is changed by less than 1\%, {indicating no major structural changes in the Pt,}  which suggests that $\theta_{\text{SHE}}$ and $\lambda_{\text{sf}}$ remain the same throughout this treatment\cite{sagasta2016tuning, laczkowski2017large}. Knowing the product $\theta_{\text{SHE}} \lambda_{\text{sf}}$ to be 0.18 nm \cite{sanchez2014prl}, we conclude that the spin mixing conductance is improved from $g_{\uparrow\!\downarrow}=0.64\,\times 10^{18}\,$m$^{-2}$ to 1.90\,$\times10^{18}\,$m$^{-2}$. The enhanced value is comparable to the ones obtained after Ar$^+$-ion milling process \cite{velez2016competing} and "piranha" etch \cite{jungfleisch2013improvement,putter2017impact}, which means that the observed increase of $g_{\uparrow\!\downarrow}$ is more a catching up of the deficit of spin conductance probably associated with the use of evaporation rather than an overall improvement of the result from what is obtained by sputtering.\\

{Next}, we focus on spin pumping measurements using the same sample batch. The experiment is performed at a fixed frequency of 9.65\,GHz in a X-band cavity while applying a static in-plane magnetic field perpendicularly to the Pt nanostrip ($x$-axis). Conversely the rf magnetic field $h_\text{rf}$ is
applied along the Pt nanostrip ($y$-axis). The rf power is fixed at 5$\,$mW ($\mu_0 h_\text{rf}=2$\,$\mu$T) to maximize the SP signal while minimizing non-linear effects such as distorsion of the lineshape (foldover effect) \cite{fetisov1999nonlinear,gui2009foldover}. At the ferromagnetic resonance condition, a flow of angular momentum from the YIG relaxes into the Pt \cite{tserkovnyak2002spin}. 

\begin{figure}
\begin{center}
\includegraphics[width=8cm,height=5cm]{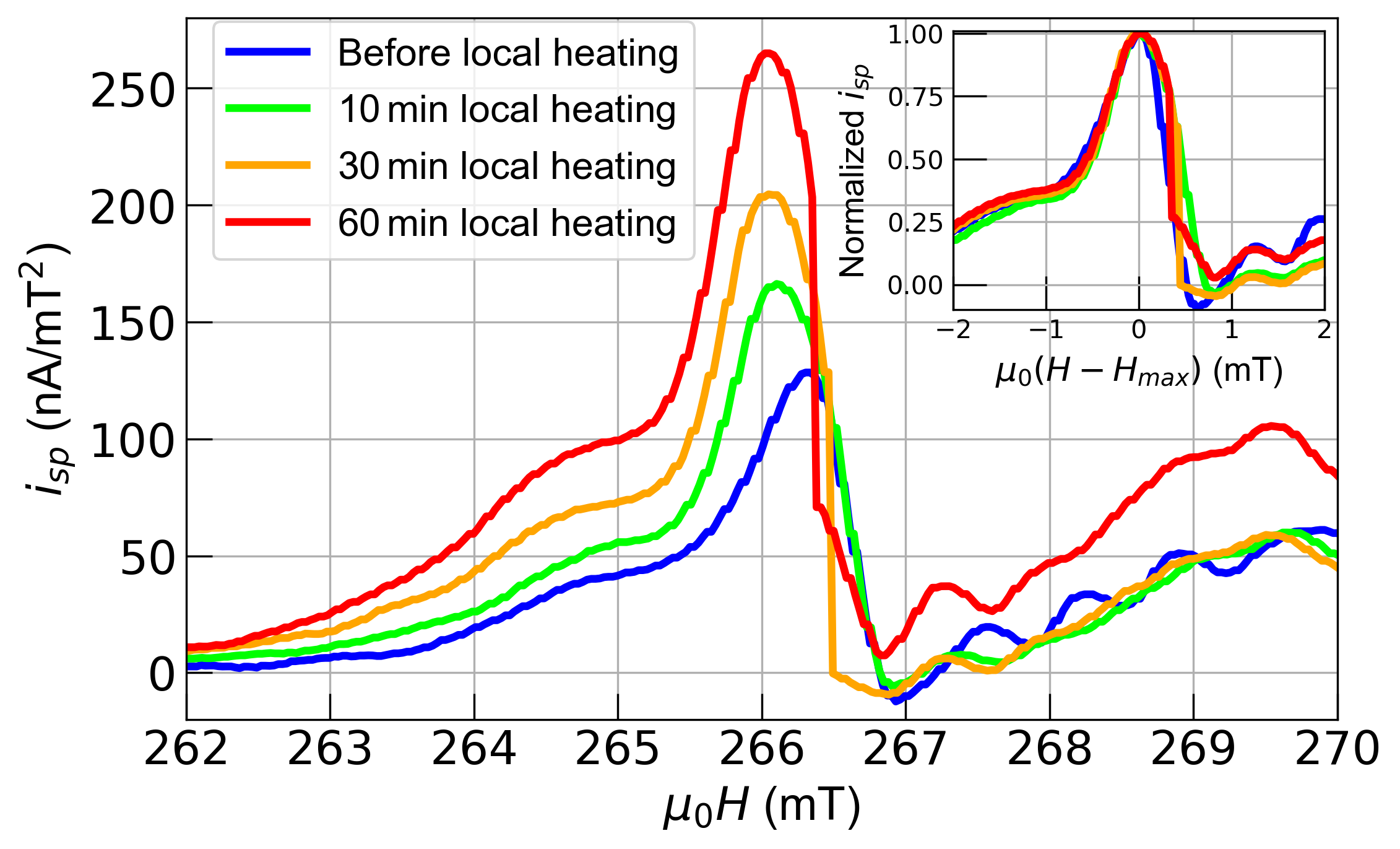}
\end{center}
\label{Figure_Sp}
\caption{ Spin pumping spectra at different annealing step by injecting a current density $J_\text{max}=1.2\times10^{12}$ A/m$^{2}$ into the Pt nanostrip for three different annealing time. 
The inset shows the normalized spin pumping spectra at each annealing time.}
\label{Fig2}
\end{figure}

The normalized spin-pumping spectra $i_\text{sp}$ are obtained by dividing the generated inverse spin Hall effect (ISHE) voltage by both $h_\text{rf}^2$ and the Pt resistance $R_0$, \textit{i.e.} $i_\text{sp} = V_\text{ISHE} / (R_0\,h_\text{rf}^2)$. 
The shape of the spectral line does not follow a simple Lorentzian, which is attributed to inhomogeneous broadening.  The main peak of the spectrum is {identified as the Kittel mode (uniform precession)} and we have {measured its amplitude} to estimate the efficiency of spin transmission {through} the interface. After characterizing the pristine YIG$|$Pt interface, we performed a local annealing by applying {pulse} current density of $J_\text{max}$=1.2$\times 10^{12}$ A/m$^{2}$ for various durations (10, 30 and 60 minutes) following the same procedure as before. The effect of annealing on the spectra are shown in Fig.\ref{Fig2}. Similar to the SMR measurement, Joule annealing increases the spin pumping signal. 
From the amplitude of the main peak of the spectra and using the model in ref.\cite{tserkovnyak2002spin}, one can estimate the enhancement of $g_{\uparrow\!\downarrow}$ from 0.60\,$\times 10^{18}\,$m$^{-2}$ to 1.23\,$\times 10^{18}\,$m$^{-2}$, compatible with our previous SMR estimation.

The interesting feature of this experiment lies in the measurement of the full width at half maximum (FWHM) of the main peak. As the amplitude of the peak become larger with the annealing time, FWHM remains constant over the whole annealing process (see the inset of Fig.\ref{Fig2}). The modulation of FWHM can be in general attributed to the extra damping induced by the coupling between YIG and Pt\cite{Heinrich2011,mosendz2010detection}. The constant FWHM reveals that the linewidth is mostly controled by the bare YIG film beside the nanometric Pt nanostrip, rather than the sole relaxation of precession dynamics at the YIG$|$Pt interface. This suggests that the additional relaxation channel provided by the adjacent metallic nanostrip is a weak perturbation of the overall relaxation of the extended YIG thin film underneath. Since this finding differs from what is observed when the adjacent Pt covers the whole YIG film \cite{beaulieu2018temperature}, we attribute the difference to finite size effects. For nanostructured Pt electrodes, the additional relaxation channel of the magnons in the YIG provided by the adjacent metal becomes weak when the lateral size of the Pt is smaller than the magnon wavelength. We emphasize that such scenario would then mostly concern the long wavelength magnons, such as those excited by the cavity.


\begin{figure}
\begin{center}
\includegraphics[width=9cm,height=12cm]{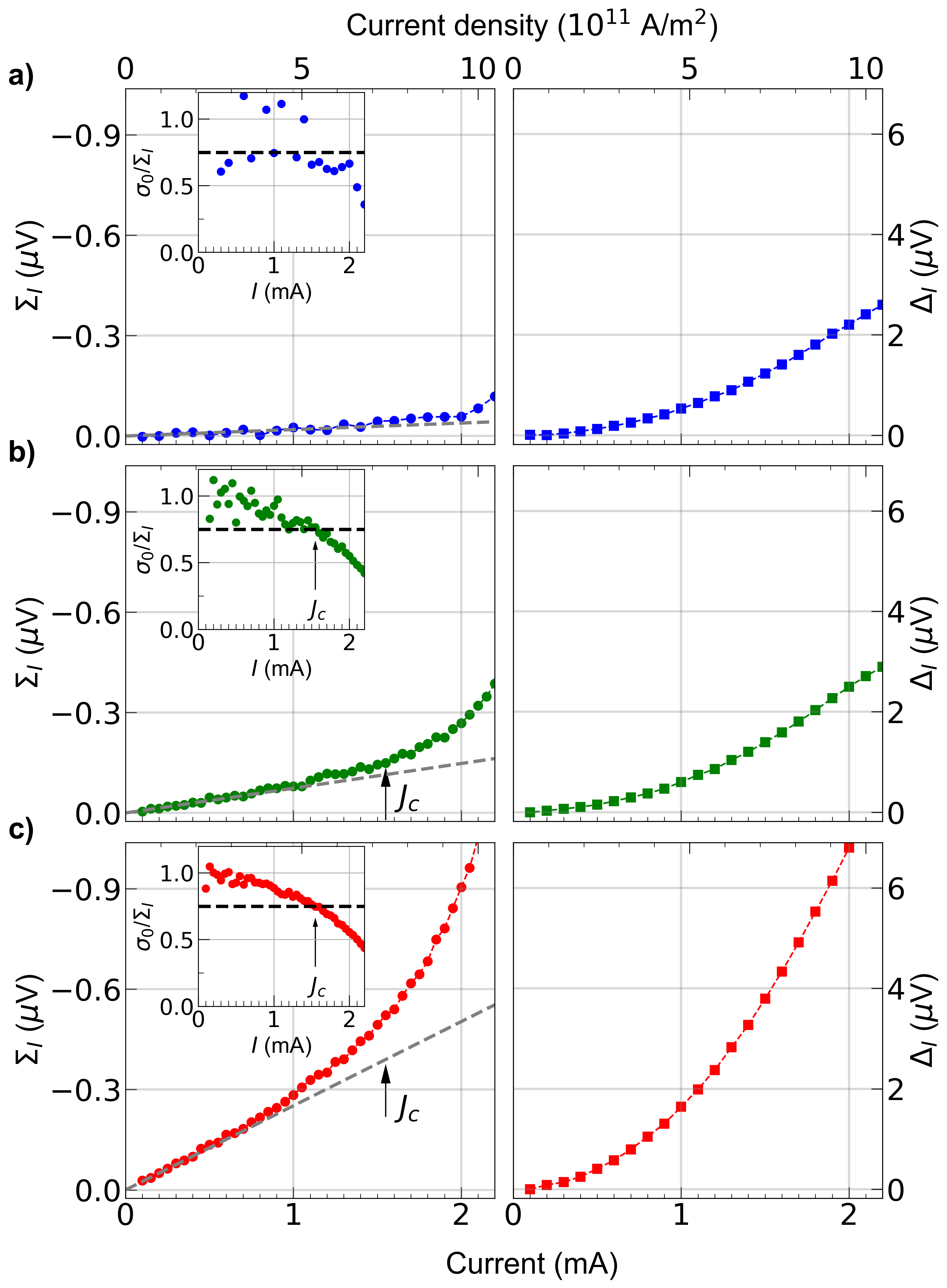}
\end{center}
\caption{Non-local spin signals $\Sigma_I$ (left) and $\Delta_I$ (right) as a function of injected current for different annealing steps. The inset of the left panel displays the current variation of $\sigma_0/\Sigma_I$, where $\sigma_0$ is the initial slope of the spin current increase (see gray dashed line). The intercept with the 0.75 level is a landmark that defines $J_c$ \cite{thiery2018nonlinear}. The (a) panel represents non-local spin signals measured directly after the nanofabrication. The (b) panel shows the result when the injector is annealed. The (c) panel shows the result when both injector and detector nanostrip are annealed.}
\label{Fig3}
\end{figure}

Finally we study the impact of local Joule annealing on the non-local spin transport properties. In these samples, the second Pt nanostrip (Pt$_2$), placed 2\,$\mu$m away from the first nanostrip (Pt$_1$), is used as a detector of the spin current (see Fig.\ref{Fig1}a). 
When a charge current is sent to the Pt$_1$ nanostrip (injector), a spin accumulation is generated at the YIG$|$Pt interface due to the SHE and its angular momentum is transferred from Pt to YIG \cite{zhang2012magnon,zhang2012spin}. The angular momentum is then carried in YIG by magnons 
, which are then detected via the ISHE voltage at the Pt$_2$ nanostrip (detector). 
Again, the non-local spin transport {is} probed by applying pulses {of electrical} current in the injector while simultaneously {monitoring} the non-local voltage $V_{\varphi}$ on the detector {as a function of the  angle $\varphi$}. {Similar to the SMR measurement, we use the background-subtracted voltage $\delta V_\varphi = V_\varphi - V_{\|}$  to extract the spin contribution \cite{thiery2018nonlinear}. The background is again measured by applying the external field $H_0$ parallel to the nanostrip direction ($y$-axis). We distinguish SSE from SOT by defining two quantities based on the $yz$ mirror symmetry \cite{thiery2018nonlinear}:  $\Sigma_{\varphi,I},\Delta_{\varphi,I}  \equiv \left ( \delta V_{\varphi,I} \pm \delta V_{\overline \varphi ,I} \right ) /2$,  where $\overline \varphi = \pi-\varphi $.} 
 
 Previously we have reported \cite{thiery2018nonlinear} that in 18\,nm thick YIG film, injecting {electrical current above a cross-over} threshold of the order of $J_\text{c}$=6.0$\times$10$^{11}$A/m$^2$ is sufficient to excite low energy magnetostatic magnons (in the GHz range). This phenomenon is characterized by the emergence a non-linear spin conduction as a function of the applied current. {In the following, we want to use this incident change of the interface transmission to investigate the influence of non-equilibrium spin accumulation at the YIG$|$Pt interface on the non-linear properties.}

Let us first consider in Fig.\ref{Fig3}a) the {pristine} state where neither interface of YIG$|$Pt at the injector nor the detector {has been} treated by {Joule} annealing. In the top panel a) of Fig.\ref{Fig3}, we display the non-local $\Sigma_{\varphi=0,I}$ and $\Delta_{\varphi=0,I}$ as a function of the applied current in the injector. 
In this configuration Pt$_1$ is connected to the current source while we record $V_2$ the voltage drop across Pt$_2$ (cf. Fig.\ref{Fig1}a).
It can be seen that the $\Delta_{I}$ shows a quadratic rise due to its thermal origin while the $\Sigma_I$ {seems to evolve} quasi-linearly with current $I$ on the range explore. The non-linear properties are analyzed by plotting in the inset of Fig.\ref{Fig3}a) the current dependence of $\sigma_0/\Sigma_I$, where $\sigma_0 =\left . (\partial \Sigma_I/ \partial I) \right |_{I=0}$ is the slope of a linear regression through the $\Sigma_I$ data measured at $I<0.5$\,mA (see gray dashed lines). We define $J_c$ as the intercept with the 75\% decrease \cite{thiery2018nonlinear} \footnote{In a single mode model, $\sigma_0/\Sigma_I$ follows a parabola, which intercepts the abscissa at the threshold current density for damping compensation, $J_\text{th}$. In this picture, the 75\% landmark corresponds to half this value $J_c=J_\text{th}/2$.}. We report on the main graph the estimate of $I_c$, which is here not precise due to the low spin conductivity of the device. Next, we perform Joule annealing of the injector (Pt$_1$) for 60 minutes with the same procedure as before. 
The impact on the non-local signals can be seen in the second panel b). We observe that the $\Sigma_I$ is now 3 times larger, indicating that the excited magnon density in YIG is enhanced due to the higher spin transmission at the injector. Remarkably, now we are able to distinguish clearly on the figure at $J_\text{c}=7\times 10 ^{11}$A/m$^2$, the cross-over threshold from a linear to a non-linear spin transport regime, which is the signature of the participation of low energy magnetostatic magnons to spin transport. Such magnons are in principle solely excited by SOT exerted at the interface between YIG and the injector. 
We also note that the  $\Delta_I$-signal produced by thermally generated magnons remains unchanged. This is expected because the spin conversion of SSE signal occurs only at the detector Pt$_2$ nanostrip (which is not annealed for this moment) whereas the injector nanostrip only plays the role of a heater. 
The last step, shown in panel c), both injector and detector are annealed. The $\Delta_I$ is enhanced by a factor of 3 due to the higher spin conversion of the probing interface.  As can be seen there too is that the $\Sigma_I$ is now 9 times larger than the reference (panel a). It follows the fact that injection and detection of magnons at each YIG$|$Pt interface are now 3 times more effective, leading to a factor of 9 increase of the SOT signals. Nonetheless, the crossover current density $J_c$  \cite{thiery2018nonlinear} is not affected by the annealing and it occurs at the same value for both cases (b) and (c). It supports the observation in SP experiments, where local annealing hasn't induced additional broadening the linewidth (inset of Fig.\ref{Fig2}). It also highlights the striking difference of out-of-equilibrium behavior between closed (nano-pillar) and open (extended films) magnetic geometries.
A possible explanation compatible with the observed behavior could be an improved wettability of the Pt on YIG after local Joule annealing {\footnote{For the sake of completeness, we have also annealed the whole sample in an oven at the same temperature during an hour. No significant changes of the interfacial spin conductance were observed, indicating the importance of performing local Joule annealing.}}. The effect could be parametrized by introducing an additional transmission coefficient $0 < {\mathcal{T}} \leqslant 1$ to the spin transparency of the interface. This coefficient represents for example the effective contact area ratio between the YIG and the Pt. This transmission alters the effective spin conductance $g_{\uparrow\!\downarrow}= {\mathcal{T}} G_{\uparrow\!\downarrow}$, while retaining constant $G_{\uparrow\!\downarrow}$ the intrinsic spin mixing conductance inferred from the interfacial spin pumping contribution to the linewidth measured in YIG films completely covered by Pt. Since the enhancement of a factor of 3 is consistently observed in all the devices, we believe that the change of contact area must have a geometrical origin probably linked to the nanofabrication process.  Because the temperature does not exceed 550$\,$K {during the heating pulses}, it is unlikely that the chemical and structural quality of the YIG surface is affected by this treatment \cite{rao2018liquid}. Thus we expect the local Joule annealing to affect only the interface between Pt nanostrip and the YIG layer \cite{papaioannou2013optimizing} and could result in a larger coverage of Pt onto the YIG surface. This improves the number of spin transmission channels at the interface which posses similar spin mixing conductance. Through this mechanism the emission, reflection or absorption of the spin current are enhanced. \\

In summary, we consistently observed enhancement of the spin-induced voltage at the YIG$|$Pt interface after local Joule annealing of evaporated Pt nanostrips. This enhancement possibly occurred through a higher coverage of Pt on the YIG surface, increasing the number of spin transmission channels available. This results can also explain the large difference in $G_{\uparrow\!\downarrow}$, $\theta_{\text{SHE}}$ and $\lambda_{\text{sf}}$ in works performed via different deposition and measurement methods  \cite{jungfleisch2011temporal,vlietstra2013exchange,liu2011review}. Additionally the spin pumping measurements and non-local magnon transport measurements showed that an enhancement of the spin transmission does not involve necessarily  an increase  or reduction of the cross-over threshold current to excite subthermal magnons, pointing out the important role of the {bare} YIG film away from the Pt nanostrip in the relaxation of {pure spin current transport}.  




\bibliography{bib_ECA}

\end{document}